\documentclass[conference]{IEEEtran}
\usepackage{lineno} 
\IEEEoverridecommandlockouts
\usepackage[table,xcdraw]{xcolor} 
\usepackage{cite}
\usepackage{amsmath,amssymb,amsfonts}
\usepackage{algorithmic}
\usepackage{textcomp}
\usepackage{ifthen}
\usepackage{enumitem}
\usepackage{pifont}
\usepackage{rotating}
\usepackage{ulem}  
\usepackage{balance}  
\usepackage{lipsum}
\usepackage{hyperref}
\usepackage{xspace}
\usepackage{enumitem}
\usepackage{float}
\usepackage{tcolorbox}
\usepackage{graphicx}    
\usepackage{subcaption}  
\usepackage{booktabs}  
\usepackage{placeins} 
\usepackage{array}     
\usepackage{setspace} 
\usepackage{wrapfig}
\usepackage{hyperref}
\hypersetup{
    colorlinks=true,
    linkcolor=blue,
    filecolor=magenta,      
    urlcolor=cyan,
    pdftitle={Overleaf Example},
    pdfpagemode=FullScreen,
    }
\usepackage{amsmath} 
\usepackage{stfloats}  
\usepackage{array}
\usepackage{multirow}
\tcbuselibrary{listingsutf8} 
\usepackage{soul}
\usepackage{lscape} 
\usepackage{tikz} 
\usepackage{tabularx} 
\usepackage{caption}  
\usepackage{array}    
\usepackage{multicol} 
\usepackage{balance}  
\usepackage{float} 
\definecolor{BlueGray}{RGB}{182,205, 216}
\definecolor{DarkBlueGray}{RGB}{63, 82, 92}
\definecolor{GreenGray}{RGB}{77,107,83}
\definecolor{safagreen}{HTML}{659447}
\newcolumntype{L}[1]{>{\raggedright\let\newline\\\arraybackslash}p{#1}} 
\newcolumntype{C}[1]{>{\centering\let\newline\\\arraybackslash}p{#1}} 
\newcolumntype{R}[1]{>{\raggedleft\let\newline\\\arraybackslash}p{#1}} 

\usepackage{seqsplit}
\newcommand{\bcode}[1]{\texttt{\seqsplit{#1}}}

\def\BibTeX{{\rm B\kern-.05em{\sc i\kern-.025em b}\kern-.08em
    T\kern-.1667em\lower.7ex\hbox{E}\kern-.125emX}}
    
\newboolean{showcomments}
\setboolean{showcomments}{true} 
\ifthenelse{\boolean{showcomments}}
  {
   
  }

\newtcolorbox{prompt}{ colback=lightgray!5!white, colframe=gray!75!black, left=1mm, right=1mm, top=.5mm, bottom=.5mm}

\newcolumntype{L}[1]{>{\raggedright\let\newline\\\arraybackslash }p{#1}}  

\definecolor{lightred}{HTML}{e2cbd5} 
\definecolor{lightviolet}{HTML}{bdaac0}
\definecolor{lightblue}{HTML}{dbebfa}
\definecolor{grayblue}{HTML}{bdc8e5}
\definecolor{lightpurple}{HTML}{b9a2d0}
\definecolor{lightorange}{HTML}{e9c899}
\definecolor{lightyellow}{HTML}{ebd67a}
\definecolor{lightgreen}{HTML}{d1e8c7}
\definecolor{lightteal}{HTML}{90c8d7}
\definecolor{lavendar}{HTML}{cac6f9}
\definecolor{lightred1}{HTML}{e7bdcf} 
\definecolor{lightred2}{HTML}{c29ead} 

\newcommand{\promptbox}[2][]{%
    \vspace{0.5em} 
    \noindent
    \fcolorbox{gray!50}{white!95!gray}{%
        \parbox{0.96\linewidth}{%
            \raggedright             
            \color{black}            
            \small                   
            \ifx&#1&%
            \else
            \textbf{#1}\\[0.5em]     
            \fi
            {\rmfamily #2}           
        }%
    }%
    \vspace{0.5em} 
}

\begin{document}

\title{Domain-Specific Persona Creation for Autonomous Sytems}
\title{Adaptive Personas as Advocates for Shaping  Requirements @ Runtime}
\title{Dynamic Personas as Advocates for Shaping Requirements @ Runtime} 

\title{From Design-Time to Runtime: Adaptive Personas for Requirements Engineering}
\title{Beyond Static Specs: Advocate Personas for Adaptive Runtime in Mission-Critical Autonomy}
\title{Living Requirements: Bridging Safety, Ethics, and Compliance Through Runtime Personas}
\title{Persona-Centric Autonomy: Linking Design-Time Intent with Runtime Realities}
\title{Runtime Advocates: A Persona-Driven Framework for Dynamic Requirements in Autonomy}
\title{Runtime Advocates: A Persona-Driven Framework for Guiding Requirements Evolution at Runtime}
\title{Runtime Advocates: A Persona-Driven Framework for Requirements@Runtime Decision Support}

\author{\IEEEauthorblockN{Demetrius Hernandez, Jane Cleland-Huang}
\IEEEauthorblockA{\textit{Computer Science and Engineering} \\
\textit{University of Notre Dame}\\
Notre Dame, IN, USA\\
\{dhernan7, janehuang\}@nd.edu}
}

\maketitle


\begin{abstract}
Complex systems, such as small Uncrewed Aerial Systems (sUAS) swarms dispatched for emergency response, often require dynamic reconfiguration at runtime under the supervision of human operators. This introduces human-on-the-loop requirements, where evolving needs shape ongoing system functionality and behaviors. While traditional personas support upfront, static requirements elicitation, we propose a persona-based advocate framework for runtime requirements engineering to provide ethically informed, safety-driven, and regulatory-aware decision support. Our approach extends standard personas into event-driven personas. When triggered by events such as adverse environmental conditions, evolving mission state, or operational constraints, the framework updates the sUAS operator's view of the personas, ensuring relevance to current conditions. We create three key advocate personas, namely Safety Controller, Ethical Governor, and Regulatory Auditor, to manage trade-offs among risk, ethical considerations, and regulatory compliance. We perform a proof-of-concept validation in an emergency response scenario using sUAS, showing how our advocate personas provide context-aware guidance grounded in safety, regulatory, and ethical constraints. By evolving static, design-time personas into adaptive, event-driven advocates, the framework surfaces mission-critical runtime requirements in response to changing conditions. These requirements shape operator decisions in real time, aligning actions with the operational demands of the moment.

\end{abstract} 


\begin{IEEEkeywords}
runtime requirements engineering, dynamic personas, event-driven architecture, human-on-the-loop, safety-critical-systems, uncrewed aerial systems (sUAS), emergency response
\end{IEEEkeywords}

\section{Introduction}

In mission-critical systems, such as emergency response with small Uncrewed Aerial Systems (sUAS), operational requirements often shift in real time \cite{al2023configuring,cleland2022extending}. Changes in weather, resource constraints, or mission goals can necessitate immediate decisions, such as recalibration of flight parameters, data collection, and autonomy levels \cite{shakhatreh2019unmanned}. To remain effective and safe in such settings, both human operators and autonomous agents must move beyond fixed, design-time requirements. Instead, they must interpret and act upon evolving, context-aware guidance at runtime.

While traditional requirements engineering for human-on-the-loop (HOTL) systems does consider operational contexts, these requirements are typically specified with the primary goal of guiding system design. They capture both functional and non-functional requirements, such as safety constraints, user interface considerations, and performance criteria, intended to shape how the system is built and validated before deployment. However, in highly dynamic environments, additional requirements often surface at runtime, influenced by emergent events, shifting autonomy levels, and human–machine interactions \cite{abeywickrama2023specifying}. This variation necessitates real-time interpretation and action that go beyond what was anticipated during design. As a result, relying solely on design-time specifications can leave operators without the necessary guidance to address unfolding mission complexities.




\begin{figure}[t]
    \centering
    \setlength\fboxsep{2pt}    
    \setlength\fboxrule{0.5pt} 
    \fcolorbox{lightgray}{white}{%
        \includegraphics[width=\linewidth]{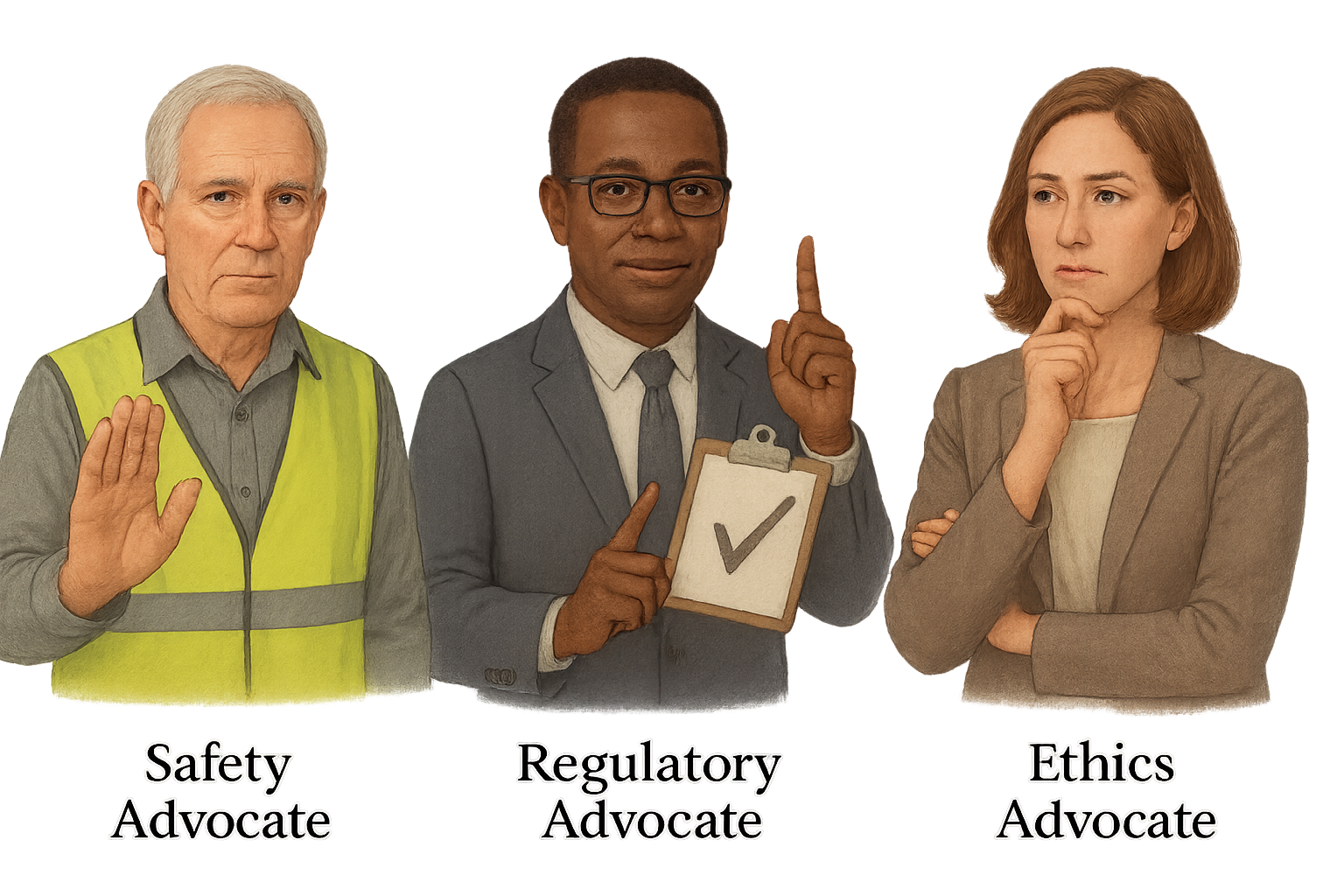}%
    }
    \caption{RAVEN leverages domain-specific Safety, Regulatory, and Ethics personas to generate contextualized runtime requirements that guide runtime decision-making by Human-on-the-loop operators. }
    \label{fig:advocate-view}
\end{figure}

To address runtime complexities in sUAS missions, we introduce RAVEN (\textbf{R}untime \textbf{A}dvocate \textbf{V}iews for \textbf{E}vent-drive\textbf{N} personas), a contextualized, just-in-time persona framework. RAVEN leverages statically defined personas to dynamically generate contextualized requirements at runtime that guide decisions associated with safety, ethics, and regulatory compliance.
To design and validate RAVEN, we adopted Wieringa's Design Science approach \cite{wieringa2014design}, which includes identifying the problem context, designing an intervention (our advocate framework), validating it against defined requirements, and evaluating its impact through empirical or simulated deployment. We conducted a proof-of-concept study involving autonomous sUAS swarms in emergency response missions. Our approach assumes that each sUAS can operate autonomously, track its own state, and respond to environmental changes detected by onboard sensors or communicated by other swarm agents, ground-based monitoring services, or human operators (e.g., \cite{al2023configuring}). Operating in safety-critical, regulated airspace requires careful attention to compliance, risk mitigation, and ethical accountability for mission success and public trust. With this context, we demonstrate how advocate personas dynamically present relevant safety, regulatory, and ethical considerations to the operator throughout the mission.

Given the characteristics of this operating environment, RAVEN uses a multi-step, in-context Large Language Model (LLM) workflow to transform structured static personas into runtime advocate personas. Further, each advocate represents a requirements-oriented advisory role. In this paper, we develop  \textit{Safety Controller}, \textit{Ethical Governor}, and \textit{Regulatory Auditor} advocates, as these represent critical concerns in the field of sUAS. However, the framework is extensible, and additional advocates, such as a Legal Advisor, an Empathetic Mediator, or a Mission Effectiveness Analyst, could be constructed to reflect other stakeholder perspectives or evolving operational needs. We ground the LLM in mission-specific world states and recognized standards (e.g., MIL-STD-882E, GDPR, FAA Part 107) via structured prompts.
This approach ensures contextual just-in-time guidance that reflects evolving operational needs. 

\begin{figure}[tbp]
\centering
\includegraphics[width=0.97\columnwidth]{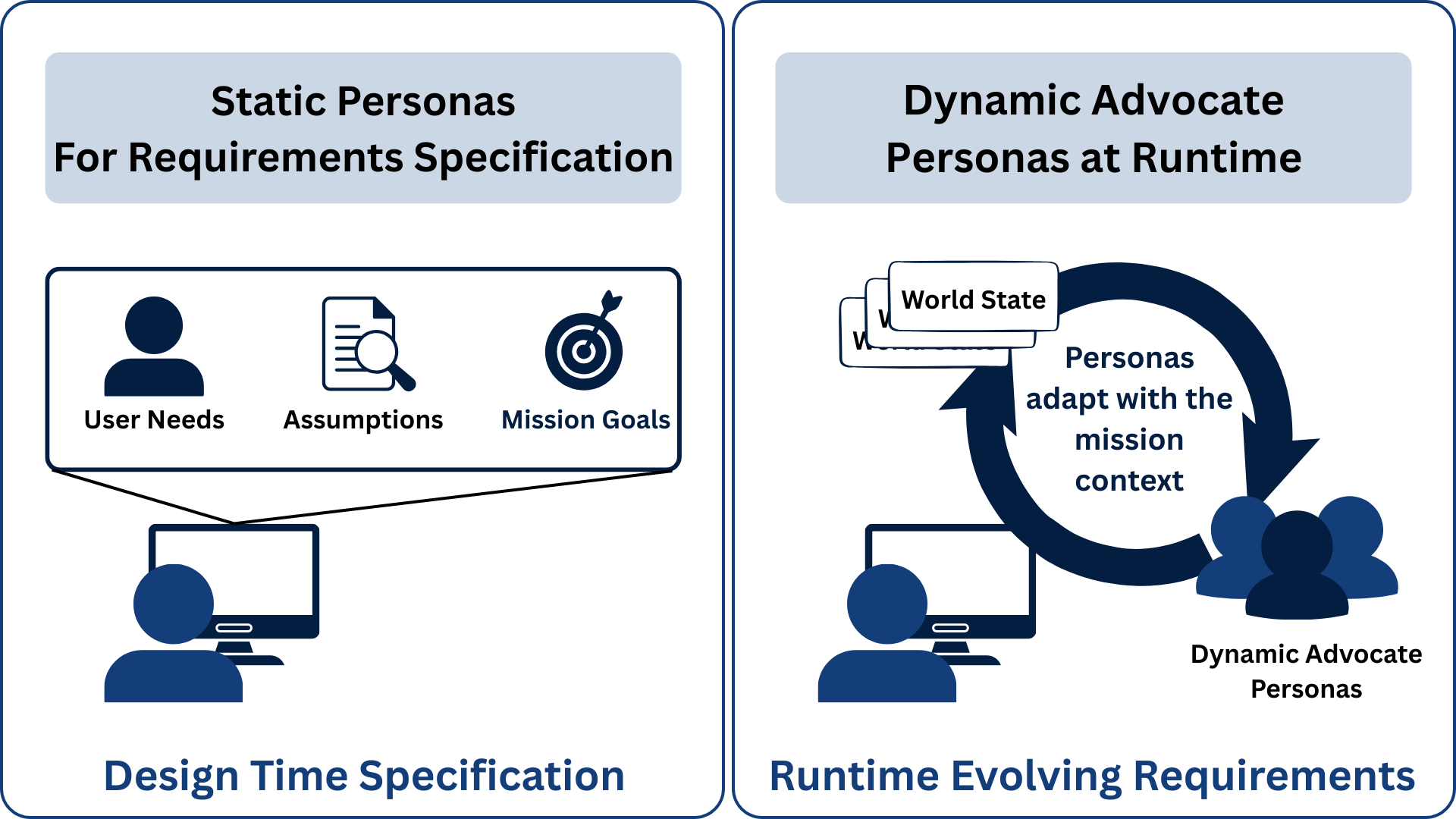}
\caption{Comparison of traditional static personas used for design-time requirements specification with dynamic advocate personas supporting runtime decision-making. Static personas rely on fixed assumptions and user needs established prior to deployment, while dynamic advocate personas adapt in response to evolving mission context, environmental conditions, and operational constraints to support informed decision making.
}
\label{fig:static_dynamic}
\end{figure}

By interleaving in-context learning with runtime persona selection and prompt orchestration, RAVEN moves beyond static design-time artifacts to enable a requirements engineering approach that adapts in real time to evolving mission conditions. An overview of the approach, illustrating the shift from static personas to runtime advocate personas, is illustrated in Figure~\ref{fig:static_dynamic}.

\section{Related Work}
Personas, a user-centered design technique, help software teams empathize with key stakeholder archetypes. However, they are often defined once (e.g., at project kickoff) and seldom updated to reflect emergent situations \cite{nielsen2021understanding}.

Prior research underscores the need to accommodate evolving operational contexts \cite{nunes2018persona}, especially in safety-critical domains, yet traditional, design-time personas do not inherently capture the nuanced changes in mission goals and real-time constraints. RAVEN addresses this gap by dynamically updating stakeholder views of safety, ethical, and regulatory concerns when triggered by environmental or operational events. To situate this work, we will examine Static vs. Dynamic Personas, Runtime Decision-Support, and Multi-Perspective Guidance.

\subsection{Static vs. Dynamic Personas}

Traditional personas in requirements engineering are static user archetypes, used primarily during early design phases to guide system requirements \cite{nielsen2021understanding}. While effective at capturing initial user needs, they lack the flexibility to adapt to evolving contexts \cite{sun2024persona, park2025charactergpt}. Recent work has explored dynamic personas that update in response to real-time data \cite{salminen2022use}, but these approaches remain limited \cite{ronkko2005empirical} and are rarely applied in safety-critical systems, despite evidence that personas can support nuanced decision-making, user modeling, and stakeholder communication across high-stakes domains such as healthcare, cybersecurity, and public health \cite{salminen2022use}. Our work extends this idea by using dynamic personas at runtime to reflect changing priorities, such as safety, ethics, and regulatory compliance.

\subsection{Runtime Human-on-the-Loop Decision Support}
Human-on-the-loop systems require decision-support tools that help human supervisors monitor and intervene in autonomous operations \cite{agha2022NeBula}. Prior research has emphasized the need for transparency, timely alerts, and explainable AI to maintain situational awareness \cite{endsley2023supporting}. Examples include human-machine systems where the human partner acts as a supervisor \cite{cleland2022extending} or ethical governors that enforce legal and moral constraints \cite{trentesaux2022engineering}. Building on this, our approach supports human-on-the-loop oversight through the use of adaptive personas as advocates, where each advocate focuses on highlighting real-time risks from a specific perspective.

\subsection{Safety, Ethics, and Regulatory Guidance}

Safety-critical domains often require balancing diverse stakeholder concerns \cite{cools2024trust}. Typically, this is addressed through hazard and risk analysis, leading to runtime features that ensure compliance with safety and legal standards \cite{li2024safetyanalyst}. However, runtime solutions tend to be focused upon functional compliance, maintaining situational awareness, and raising critical safety-related alerts (e.g., ``flying too close to the terrain''). Little work has addressed the need for operators to understand the broader regulatory, safety, and ethical concerns at runtime that shape their operational decisions \cite{feldman2024war}.  We address this need by using personas to frame stakeholder concerns that surface tailored alerts (e.g., ethical cautions or regulatory warnings) during operations. This multi-perspective guidance is designed to help operators make informed, accountable decisions in high-stakes environments.
\section{Approach}

\subsection{Advocate Personas}

\begin{figure}[tbp]
\centering
\includegraphics[width=1\columnwidth, trim=20 0 20 0, clip]{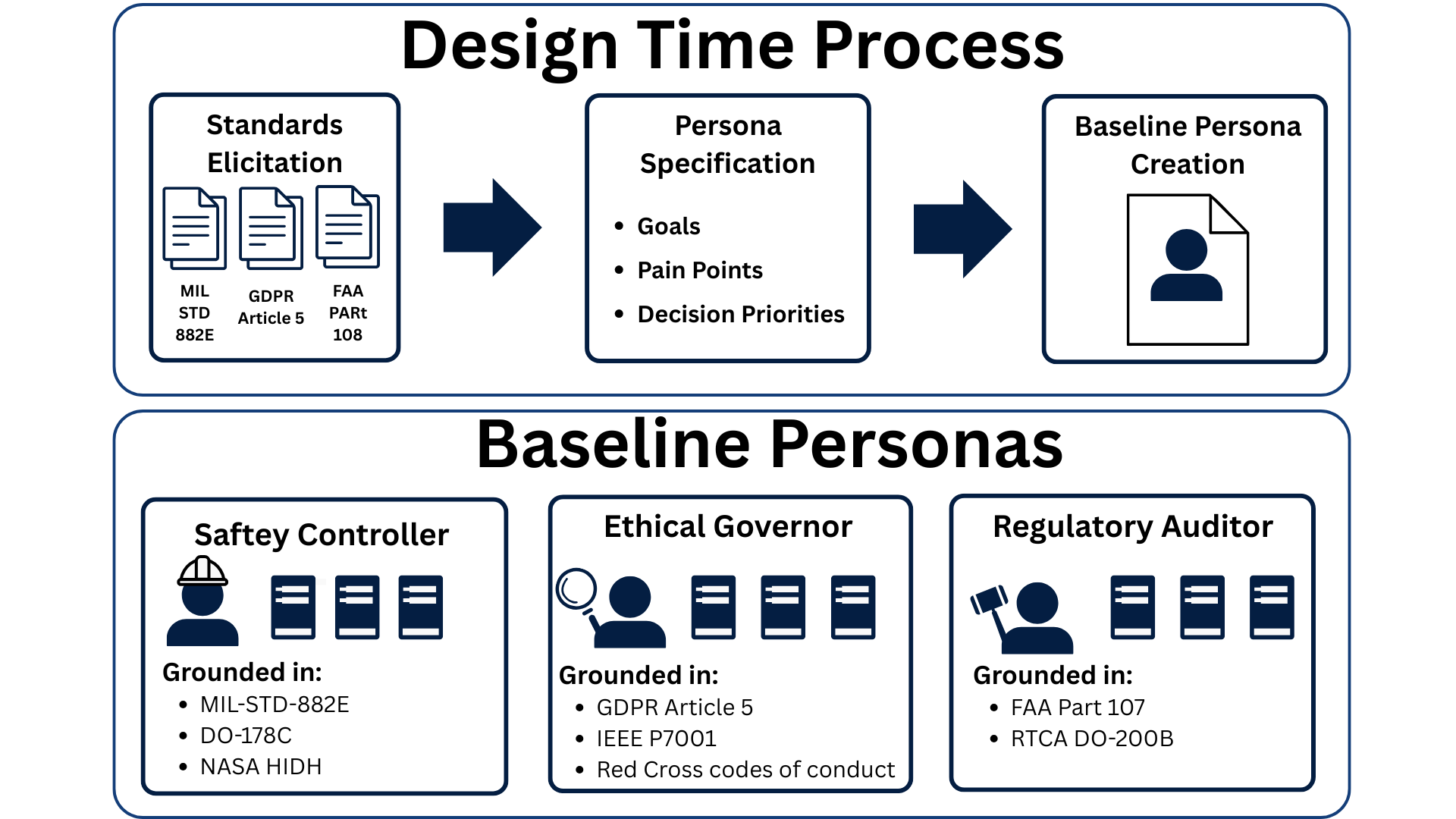}
\caption{Baseline persona generation process for RAVEN.
}
\label{fig:baseline_personas}
\end{figure}
Event-driven personas can be conceptualized as a specialized form of dynamic persona whose influence on decision support is directly and continuously shaped by significant events that occur within the operational environment, or through interactions initiated by the human operator. Our design of RAVEN, therefore, draws upon the well-established concepts of event-driven architecture (EDA) that are increasingly prevalent in modern AI and autonomous systems. The personas exhibit four key EDA characteristics: (1) asynchronous communication and updates, (2) real-time responsiveness, (3) loosely coupled modular design, and (4) context-aware processing of event data. Further, each persona is rooted in a rigorous design-time specification process that begins with the elicitation of relevant standards and guidelines to define the scope of each persona's responsibilities. From this, we derive structured representations that capture each advocate's core goals, pain points, and decision-making priorities. These structured representations constitute a baseline persona that is a static artifact that codifies each persona's knowledge boundaries and value commitments prior to its deployment (see Figure~\ref{fig:baseline_personas}).

We define three advocate personas as a Safety Controller, Ethical Governor, and Regulatory Auditor. We selected these three advocates because they represent the most commonly intersecting stakeholder concerns in safety-critical, human-on-the-loop systems like sUAS swarms.  Further, these three dimensions frequently surface in real-world deployments and often introduce trade-offs that require timely, informed decisions by human operators.

Each advocate is grounded in existing standards and guidelines. For example, the Safety Controller is grounded in MIL-STD-882E for hazard assessment \cite{jensen2022selecting}, DO-178C for software assurances \cite{dmitriev2021toward}, and NASA Human Integration Design Handbook (HIDH) for fail-safe best practices \cite{silva2019implementation}. The Ethical Governor is grounded in GDPR Article 5 on data minimization \cite{nasir2024ethical}, IEEE P7001 for transparency \cite{ieee2022ieee}, and Red Cross codes for humanitarian conduct \cite{zomignani2020aid}. Finally, the Regulatory Auditor is based on FAA Part 107 regulations for sUAS operations \cite{greenwood2020flying} and RTCA DO-200B for logging and traceability \cite{marques2017verification}, among others.

The personas respond to evolving contexts, adapting their advice to provide up-to-date situational awareness and decision support as states change and new events occur. This enables RAVEN to deliver timely, context-aware guidance to the human operator. As loosely coupled components, each persona can focus on a specific facet of the swarm's operation or a particular aspect of human-swarm interaction, providing greater flexibility and scalability in managing system complexity. As events occur, the personas actively interpret embedded contextual cues and generate actionable insights that reflect the current mission landscape.

We anchor their outputs in in-context references to recognized standards so the LLM has a better chance of producing consistent, compliance-oriented recommendations.

\subsection{Architecture Overview}

\begin{figure}[tbp]
\centering
\includegraphics[width=.8\columnwidth, height=0.45\textheight, keepaspectratio]{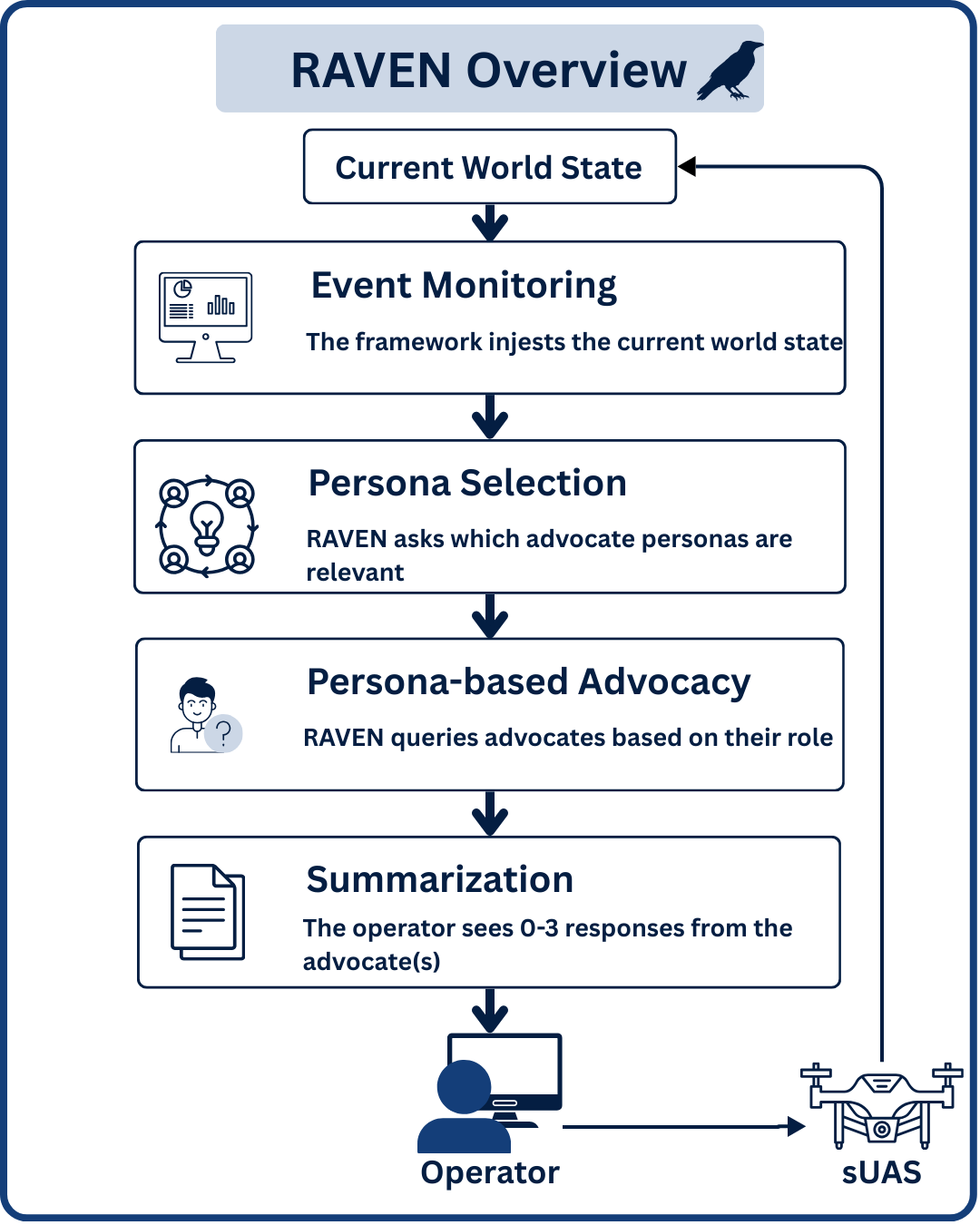}
\caption{RAVEN: Our framework where the evolving world state triggers event monitoring, initiates persona selection, and delivers 0–3 advocate responses to the operator, who then adjusts drone behavior, completing the decision-feedback cycle.
}
\label{fig:architecture_overview}
\end{figure}

Figure~\ref{fig:architecture_overview} shows an example demonstrating how RAVEN’s pipeline leverages advocate personas to provide actionable requirements at runtime. For clarity, we present abridged prompt excerpts in this section; in practice, each prompt contains additional contextual details, domain references, and structured instructions to ensure comprehensive and accurate responses. For illustrative purposes, we provide a single shared link that demonstrates the full prompt pipeline as replicated in ChatGPT. In actual operation, these prompts are generated and sent by RAVEN to the LLM via an API. See: \url{https://tinyurl.com/REnext2025}

\begin{enumerate}
    \item \textbf{Event Monitoring:} RAVEN ingests a continuously updating world state that reflects environment, system, and operator conditions. Events are defined as significant changes in state and are detected through runtime monitoring. When such an event occurs, it activates the pipeline to assess relevant advocate personas.

    \promptbox{
    \textbf{Trigger:} 

    World state updated. Current wind speed: 22 mph, forecast: WORSENING.
    
    }\vspace{6pt}
    
    \item \textbf{Persona Selection:} RAVEN uses a fixed prompt to ask which advocate personas are relevant based on the current world state. The prompt summarizes key context and explicitly asks which domains (safety, ethics, regulation) are affected. For example:

    \promptbox{
    
    \textbf{Condensed Prompt:}

    Given a 22 mph wind with a worsening forecast, which advocate persona(s) are relevant?
    
    \textbf{Examples of Expected Response:}
    
      \textit{\emph{selected\_advocates}}: ["Safety Controller"]
      
      \textit{\emph{rationale}}:
      
         Safety Controller: "Wind levels pose stability concerns; immediate attention is required."
         
         Ethical Governor: "No privacy or humanitarian conflict is evident right now."
         
         Regulatory Auditor: "No active regulatory issues triggered by wind conditions."
         
    }\vspace{6pt}
    
    \item \textbf{Persona-Based Advocacy:} For each identified advocate (e.g., the Ethical Governor if privacy or fairness is implicated), RAVEN injects domain references and relevant scenario data as in-context prompts. Then RAVEN asks advocates specific questions based on the advocate's role. The answers are grounded in the relevant standards.

    \promptbox{
    \textbf{Condensed Prompt:}
    
    Safety Controller, the wind is 22 mph with a worsening trend. Please provide immediate guidance. 
    
    \textbf{Examples of Expected Response:}
    
    1) Advise the operator to lower the altitude and reduce speed to mitigate gust-related instability.

    2) Monitor for sudden shifts in wind direction and be prepared to pause or abort the mission if stability degrades.

    3) Confirm all critical sensors remain active for real-time turbulence tracking.
    }\vspace{6pt}

    \item \textbf{Summarization:} The generated advocate-specific advice is then fed back into the LLM as a separate summarizing step. As a result, the operator is presented with a concise set of warnings or recommendations.

    \promptbox{
    \textbf{Condensed Prompt:}
    
    Summarize the key points for the operator in two sentences.
    
    \textbf{Examples of Expected Response:}
    
    1) High wind gusts require immediate adjustments to flight parameters for operational safety.
    
    2) Regularly assess changing conditions and be ready to suspend the mission if turbulence worsens.
    }
    
\end{enumerate}

In this initial proof-of-concept, no external retrieval system (RAG) was used; instead, we relied on carefully curated in-context domain references to ensure that LLM’s outputs stayed accurate and relevant.

\section{RAVEN's Iterative Design}

To iteratively validate and refine the behavior and effectiveness of RAVEN, we developed 15 scenarios designed to represent a diverse range of operational and contextual challenges within the domain of sUAS-swarming for emergency response. We refer to these scenarios as \textit{formative} tests, as they were used to iteratively refine the RAVEN infrastructure, including its static personas and runtime LLM prompts.

We started by establishing a baseline scenario featuring ideal operating parameters: clear visibility, low obstacle density, stable system status, proper flight authorizations, and no active hazards. We showed that, as expected, our advocates did not generate runtime guidelines under these normal operating conditions.  

The set of formative scenarios was then grouped into three domain-specific categories: Safety, Ethics, and Regulatory, as well as five complex cross-domain cases that simultaneously stress safety, ethics, and regulation. These included scenarios characterized by high wind conditions, low battery warnings, privacy concerns in densely populated areas, expiring flight authorizations, and mission-critical tasks near restricted airspace.

To support the iterative design and refinement process, we developed a lightweight test environment that simulated dynamic sUAS states, environmental conditions such as weather, visibility, and restricted zones, as well as human inputs. This environment was monitored by RAVEN, which assessed the evolving context, selected relevant advocate personas, and routed scenario-specific data. The advocates were responsible for generating runtime requirements tailored to the current mission context.
We inspected the outcomes for each of the formative test scenarios to identify flaws such as generic responses, inaccurate persona selection, or failure to anticipate near-term risks, and used these observations to drive targeted improvements. Improvements included refining prompt phrasing and restructuring how scenario variables were represented. Through repeated cycles of scenario testing, manual inspection of results, and revisions, we iteratively improved RAVEN's responses.

Importantly, these formative scenarios were used exclusively for pipeline development and refinement, and an additional set of scenarios was used for evaluation.
\section{Evaluation and Results}

\subsection{Experimental Design}

To evaluate the advocate personas in realistic and diverse operational contexts, we froze the prompt templates and system pipeline developed during the design process, in order to ensure consistent evaluation conditions. We elicited new test scenarios from two researchers with expertise in the sUAS domain. They were shown an example of one formative scenario and asked to generate additional scenarios that reflected normal mission challenges and plausible edge cases. Our goal was to assess whether the RAVEN framework (1) correctly identified and activated the appropriate advocate persona(s) based on the current world state, (2) generated domain-relevant and context-sensitive guidance aligned with each persona’s role, and (3) avoided conflicting or redundant recommendations in multi-persona situations. The test environment was able to accommodate all proposed test scenarios by updating the world state, which served as input into our test oracle. The oracle evaluates snapshots of the current world state as fed into RAVEN. All of the test scenarios and corresponding outcomes are presented in Table~\ref{tab:drone_advocates_actual}.

\newcolumntype{Y}{>{\raggedright\arraybackslash}X} 

\begin{table*}[t]
  \centering
  \scriptsize                              
  \setlength{\tabcolsep}{4pt}              
  \renewcommand{\arraystretch}{1.1}        
   \caption{Verbatim responses produced by the Safety Controller, Ethical Governor, and Regulatory Auditor advocates for a critical event in three test scenarios. The Safety Controller and Ethical Governor each responded in two scenarios, while the regulatory auditor offered guidance in all three cases.}
  \label{tab:drone_advocates_actual}
  \begin{tabularx}{\textwidth}{%
      >{\raggedright\arraybackslash}p{2.5cm} 
      Y                                      
      Y                                      
      Y                                      
    }
    \toprule
    & \textbf{Scenario~1} 
      & \textbf{Scenario~2} 
      & \textbf{Scenario~3} \\
    \textbf{Advocate} 
      & Battery at 15\% (5\,min endurance); forced landing on neighbor’s lawn.
      & Recording video adjacent to a secure correctional facility.
      & Operating near a 380\,ft building under 12\,mph gusts and wind shear. \\
    \midrule
    \begin{minipage}[t]{2.5cm}
      \centering
      \textbf{Safety Controller}\\[4pt]
      \includegraphics[width=1.5cm]{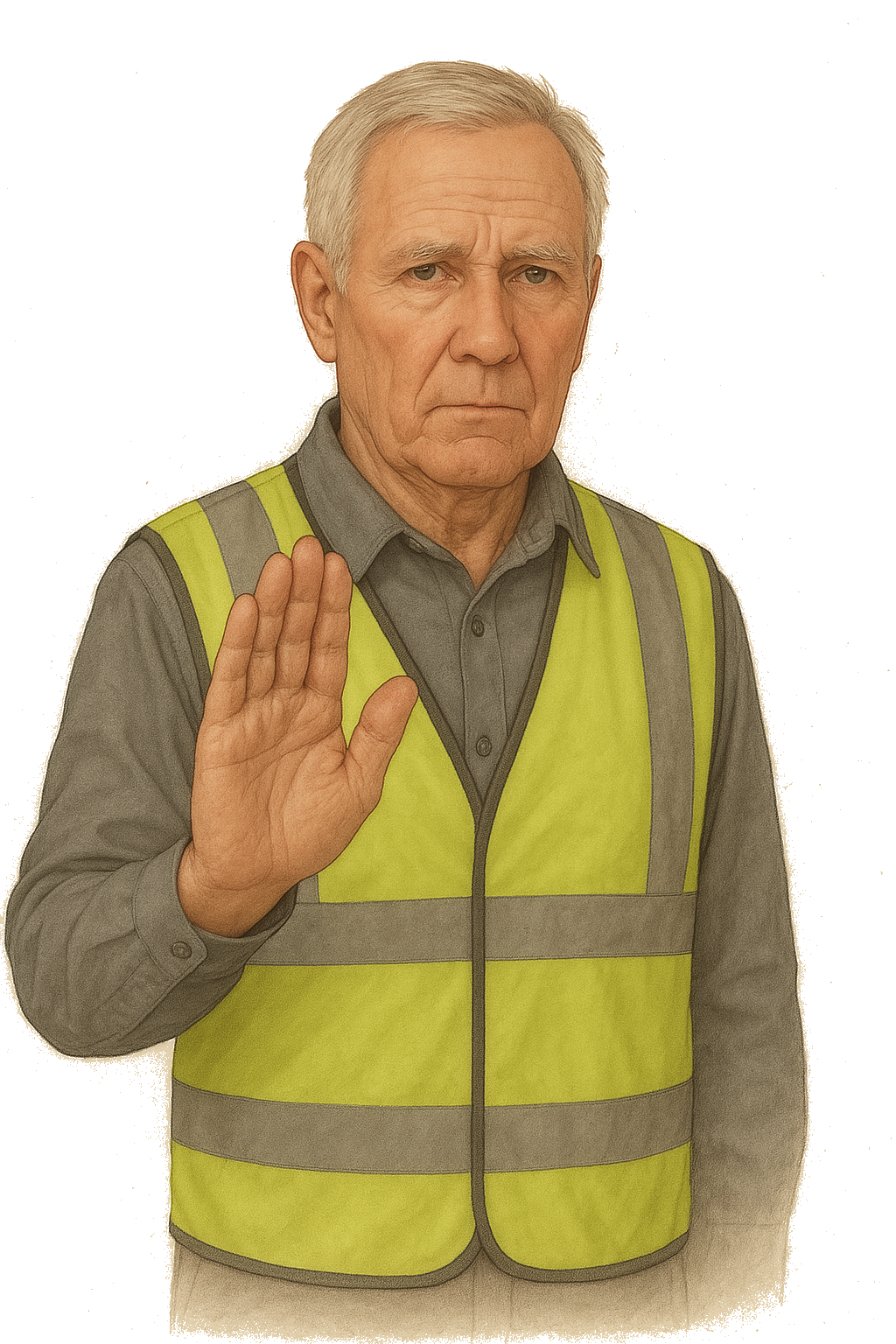}
    \end{minipage}
      & Immediate assessment of \bcode{system.status.powerLevel=15} and \bcode{system.status.estimatedEndurance=00:05:00} indicates an urgent need to expedite payload delivery or proceed with a safe landing. \newline\newline
        Due to \bcode{environment.location.populationDensity=moderate} and \bcode{regulatory.restrictedAreas.distanceMeters=10}, quickly identify or coordinate a valid landing zone to avoid property conflicts.
      & 
      & Maintain stable positioning near the 380\,ft structure by continuously monitoring pitch/roll under \bcode{windSpeedMph=12} gusts, \bcode{altitude=350}, and high obstacle density to avoid collisions. \newline\newline
        Check battery at 80\% frequently to maintain the 15\% safety margin and prevent an early return mid-inspection. \\
    \midrule
    \begin{minipage}[t]{2.5cm}
      \centering
      \textbf{Ethical Governor}\\[4pt]
      \includegraphics[width=1.7cm]{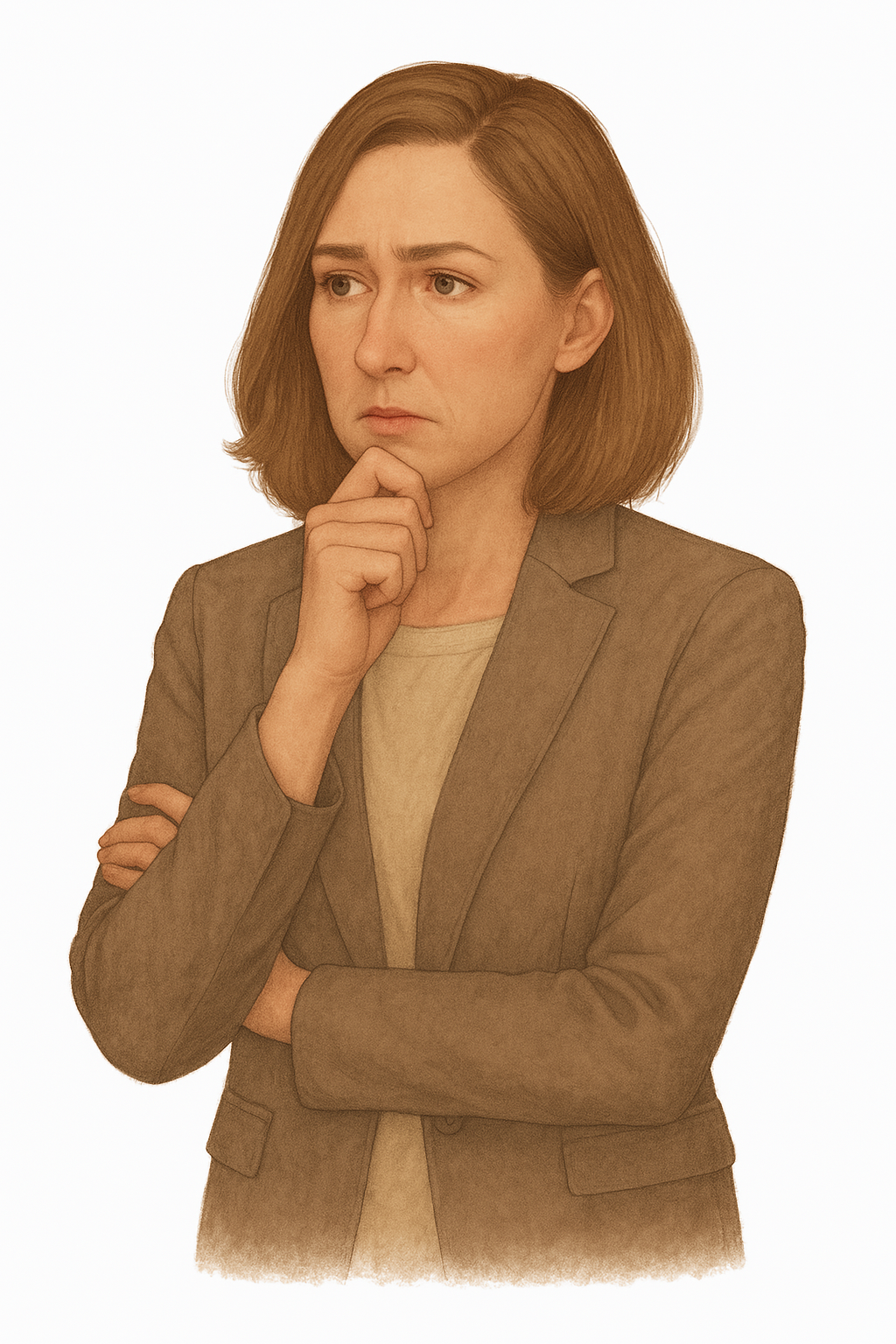}
    \end{minipage}
      & Respect immediate privacy concerns by avoiding data capture in the \bcode{'private\_property'} area (\bcode{worldState.regulatory.restrictedAreas.nearestType=private\_property}) and conforming to \bcode{'dataOperations.collectionLevel=none'}. \newline\newline
        Simultaneously remain vigilant for newly identified individuals needing aid via \bcode{'emergency\_first'} prioritization (\bcode{worldState.operationalParameters.resourceManagement.prioritizationMethod}) and real-time coordination with \bcode{'team\_alpha'} (\bcode{worldState.operationalParameters.resourceManagement.groundTeams}).
      & Adjust your flight path and camera angles to avoid restricted prison monitoring at 80\,m, as indicated by \bcode{restrictedAreas} and \bcode{missionConstraints.sensitiveAreaHandling}. \newline\newline
        Increase sensor sweeps due to moderate populationDensity and high vegetationDensity to ensure no potential survivor remains undiscovered.
      &  \\
    \midrule
    \begin{minipage}[t]{2.5cm}
      \centering
      \textbf{Regulatory Auditor}\\[4pt]
      \includegraphics[width=1.6cm]{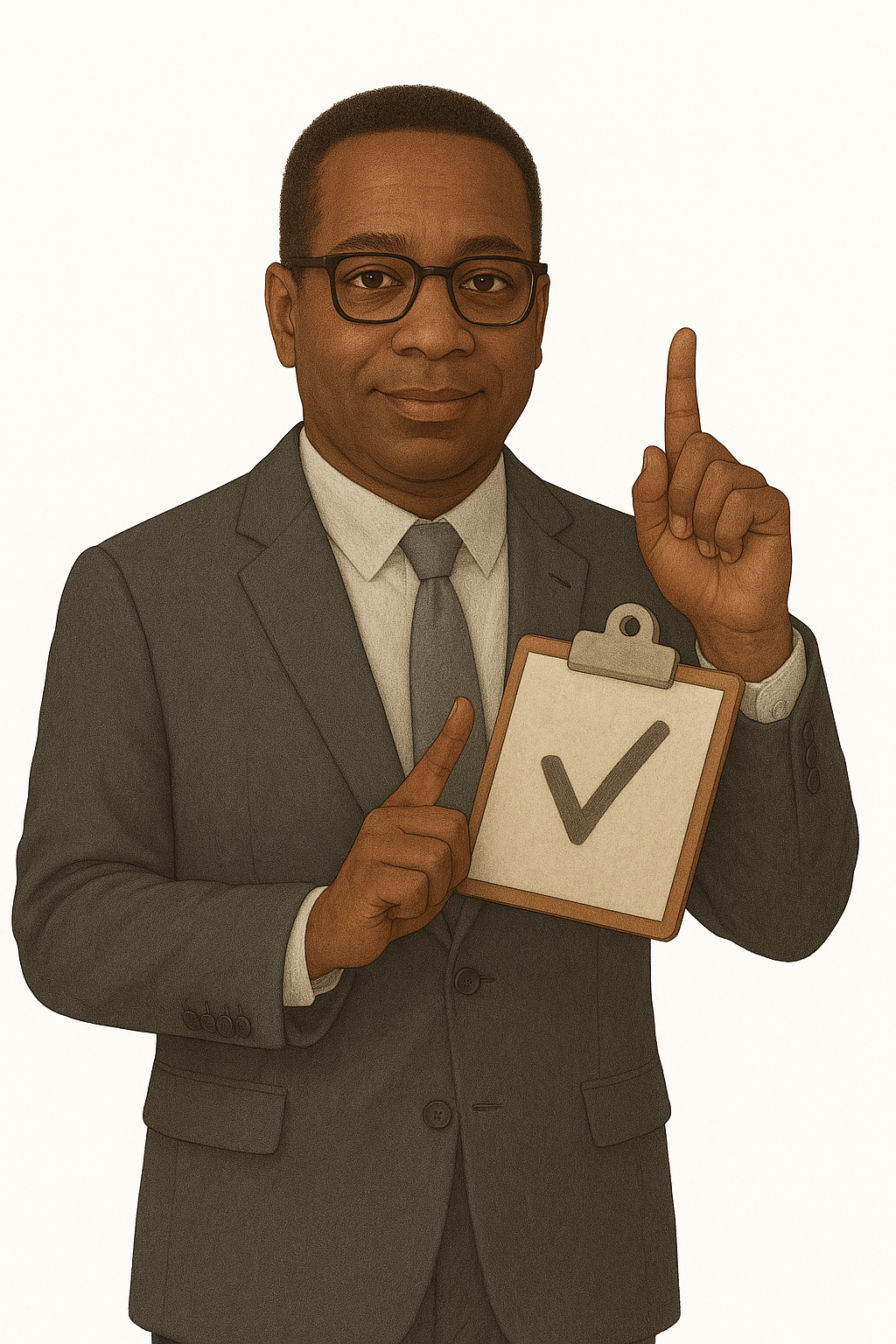}
    \end{minipage}
      & Maintain strict separation from the restricted property (\bcode{regulatory.restrictedAreas.distanceMeters=10}, \bcode{restrictedAreas.notificationRequired=true}) to avoid unauthorized incursion during the final approach. \newline\newline
        Monitor power reserves (\bcode{system.platform.status.powerLevel=15}, \bcode{missionContext.elapsedTime=00:15:00}) to ensure a safe payload delivery before battery depletion.
      & Remain strictly aware that you are currently within 80\,m of the prison (\bcode{distanceMeters=80}, \bcode{notificationRequired=true}), requiring immediate vigilance to avoid capturing restricted footage and maintain compliance. \newline\newline
        Notify authorities as needed and closely monitor flight headings and sensor usage (\bcode{heading=180}, \bcode{opticalZoom=3}) to align with ongoing regulatory constraints.
      & Confirm compliance by verifying restricted area notifications using \bcode{restrictedAreas.distanceMeters=250} and \bcode{restrictedAreas.notificationRequired=true} to ensure valid operations near the tall building. \newline\newline
        Complete your inspection before \bcode{authorizationExpires=2023-06-14T20:00:00Z}, or secure an immediate extension to remain compliant with current regulations. \\
    \bottomrule
  \end{tabularx}
 
\end{table*}













\subsection{Results}
First, we summarize how each test scenario activated specific personas, alongside each advocate’s performance.

\noindent{$\bullet$ \textit{Test Scenario 1: Low Battery, No Safe Landing.~}} This scenario activated all three advocate personas. The \emph{Safety Controller} identified the risk of in-flight power depletion; the \emph{Ethical Governor} flagged trespassing concerns on a neighbor’s property; and the \emph{Regulatory Auditor} highlighted local property restrictions.

\noindent{$\bullet$ \textit{Test Scenario 2:  Lost Person Near Prison.~}}  An overlapping field of view with an adjacent prison activated the \emph{Ethical Governor} and \emph{Regulatory Auditor}. Although no acute safety hazards were present, both advocates cautioned against unauthorized video capture.

\noindent{$\bullet$ \textit{Test Scenario 3: Tall Building and Wind Shear.~}}
Wind shear near a 380 ft structure activated the \emph{Safety Controller} and \emph{Regulatory Auditor}. The Safety Controller advised maintaining stability in gusting conditions, while the Regulatory Auditor flagged proximity and altitude compliance near the tall structure. No ethical concerns were identified.

\subsection{Discussion}
We now analyze the test outcomes through the lens of three key evaluation criteria: response relevance, scope of advocacy, and advocacy alignment.

\paragraph{Response Relevance}
\textit{Response relevance} evaluates the extent to which each advocate produces advice that directly addresses the operating domain.
Across all evaluation scenarios, RAVEN’s event-driven advocate personas consistently produced context-sensitive advisories that directly addressed the specific operational triggers. RAVEN was able to transform complex world state data into actionable runtime requirements grounded in relevant domain knowledge. This responsiveness suggests that integrating event-driven personas into runtime requirements engineering can move towards closing the gap between pre-specified design-time intentions and emergent mission realities. The results support the notion that just-in-time, role-specific guidance can support the requirements surfaced to human-on-the-loop operators in safety-critical environments.

\paragraph{Scope of Advocacy}
\textit{Scope of Advocacy} evaluates the extent to which each advocate `stays in its swimlane' providing advice related to its expertise. We observed that each advocate persona within RAVEN adhered strictly to its defined scope of responsibility, avoiding overreach into adjacent domains. Safety-related advisories remained focused on physical risk mitigation, while ethical and regulatory guidance centered on moral obligations and compliance constraints, respectively. This strict alignment with persona-specific mandates reflects a core strength of RAVEN’s architectural separation of concerns. In the context of requirements engineering, this separation enables the generation of modular and interpretable guidance streams.

\paragraph{Advocacy Alignment}

\textit{Advocacy Alignment} evaluates the extent to which advice generated by each advocate aligns, rather than conflicts, with the advice of other advocates. In multi-faceted scenarios involving simultaneous safety, ethical, and regulatory considerations, RAVEN’s advocate personas delivered coherent, non-contradictory runtime advisories. However, in one scenario, the Ethical Governor recommended disabling cameras to protect privacy, while the Regulatory Auditor advised continued recording to ensure compliance and traceability. Rather than undermining alignment, these divergences highlight RAVEN's ability to surface legitimate and competing ideas transparently. The results illustrate how persona-driven frameworks can integrate specialized lenses, such as risk, compliance, and moral considerations, without diluting one another.

\section{Threats to Validity}

As a Research Preview paper, our goal has been to present RAVEN as a novel contribution, supported by an initial proof-of-concept. However, there are several threats to validity. First, RAVEN relies partially on LLM-driven persona generation. Inherent biases in the LLM training data might lead to oversights or culturally biased assumptions in the Ethical Governor domain. To address this, future work will incorporate LangChain and Retrieval-Augmented Generation (RAG) techniques, enabling users to provide domain-specific knowledge and context to guide persona generation and reasoning.

Second, while we modeled a realistic world state for the domain of sUAS and used it to evaluate realistic scenarios, other application domains may introduce further complexity. Additionally, the evaluation used a limited set of curated test scenarios designed to exercise the pipeline’s core functionality. While these scenarios cover safety, ethical, and regulatory conditions, the set is not large enough to capture the full diversity and unpredictability of real-world missions. Future work will explore RAVEN across more diverse domains and more complex scenarios. 

Finally, we have only tested RAVEN in a test-oracle designed to simulate state changes, and have not yet conducted user studies in a high-fidelity simulation environment or in live field deployments with diverse stakeholders and roles. While the proof-of-concept has shown promising results, we cannot yet claim that it is useful to human operators. Therefore, in future work, we plan to integrate RAVEN into a sUAS framework for evaluation in simulation and real-world settings with diverse stakeholders.
\section{Conclusions and Future Work}

This paper introduced RAVEN, a novel framework that transforms traditional static personas into dynamic, event-driven advocate personas capable of supporting runtime requirements engineering in mission-critical environments.  By operationalizing the roles of Safety Controller, Ethical Governor, and Regulatory Auditor, RAVEN offers human-on-the-loop operators contextual, timely, and standards-informed guidance in domains where runtime adaptability is essential.

Through an initial evaluation in sUAS-based emergency response scenarios, we demonstrated how RAVEN interprets changing world states and provides actionable, domain-specific recommendations aligned with evolving operational demands. Our findings suggest that embedding personas within a structured LLM workflow can surface runtime requirements that would otherwise remain latent or unaddressed, enhancing situational awareness and accountability.
Further, our proof-of-concept demonstrates how RAVEN contributes toward runtime requirements engineering that is responsive, context-aware, and grounded in operational realities. 

In future work we plan to fully evaluate RAVEN in a series of user-studies conducted in a high-fidelity simulation environment and in real-world environments with end-users. We will assess the usefulness and trustworthiness of the advocates to diverse users, and will compare our current push-model (where advice is dynamically pushed to the user in a similar format to warning messages), to a pull-model (in which users can seek advice upon demand), and also to various hybrid models.
We will also extend the advocates themselves in several ways. First, by considering additional types of advocates that might be relevant across the sUAS and other domains, and second, by enhancing the accuracy and usefulness of current advocates, for example, by enhancing the pipeline with machine-learning techniques to improve filtering of recommendations, and by integrating retrieval-augmented generation (RAG) methods, richer world state representations, and potentially graphical inputs to improve the reliability and fidelity of persona prompting in dynamic environments.

\IEEEtriggeratref{13}
\bibliographystyle{IEEEtran}
\bibliography{renext}

\end{document}